\magnification1200

\rightline{KCL-MTH-02-22}
\rightline{hep-th//y0208214}

\vskip .5cm
\centerline
{\bf The Spontaneous Compactification of the Closed Bosonic String }
\vskip 1cm
\centerline{  Peter West }
\vskip .5cm
\centerline{Department of Mathematics}
\centerline{King's College, London, UK}

\leftline{\sl Abstract}
\noindent 
We argue that the higher space-time derivative terms
that occur in   closed bosonic string field
theory are reminiscent of those found in the  Landau theory of
liquid-crystal transitions. We examine the lowest level  approximation of
the closed bosonic string and find evidence for the existence of a new
vacuum that  spontaneously breaks   Lorentz and
translation symmetries. This effect can be  interpreted as   a
spontaneous compactification of the theory. 
 
\vskip .5cm

\vfill

\vskip 1cm
email:  pwest@mth.kcl.ac.uk

\eject


String field theories differ from more conventional  
 field theories  in 
that  they contain an infinite number of fields. However, they also  
possess 
another  unusual feature, namely they contain 
 interactions  terms that involve an infinite number of space-time
derivatives. This feature can be thought of as a consequence of the
extended nature of the string and the causal nature of its interactions. 
In  light-cone string field theory [1], or the  gauge covariant
string field theory  of references [2,3], strings interact by  joining, 
or alternatively splitting,
at their end points. Clearly,  this interaction is
causal in that it occurs at a point in space-time. It can be described by
  overlap conditions that simply state, using appropriate delta
functions, that the strings have joined in this way. When these
interactions are formulated in terms of the infinite set of component
fields of the string one finds that the interactions have an 
infinite
number of space-time derivatives. Although the causal nature of the
interaction is less apparent in the most popular open string field theory
[4], which is based on strings joining at their mid points, the
interactions also contain an infinite number of space-time derivatives. 
\par
From early on it was realised that the  open and closed bosonic strings in
twenty six dimensions  possessed tachyons. Often the appearance of these 
tachyons  has been  cited as a reason to reject
these theories from serious consideration, although this same argument
was never deployed to argue against  the presence of a tachyon in  
the standard model  of particle physics. In fact, since the earliest
stages of the development of  string theory it has been realised that the
tachyon  could signal  an instability in the vacuum state. To
demonstrate such a  phenomenon required a non-perturbative 
formulation of these string theories,  such as  string field theories.
However, as these formulations contain an infinite number of component
fields, the calculation to demonstrate such a mechanism looked difficult.
The first clear steps to demonstrate the existence of a new vacuum   in
the open bosonic string came in reference [5] where a systematic
approximation procedure, namely level truncation,  was used.  Convincing
evidence for a new vacuum was found and it was shown, as a consequence
of the higher derivative terms, that many of the particles in the original
theory seemed to be eliminated as they did not have poles at the new
minimum [5]. 
\par 
More recently, it has been argued that this new vacuum has a natural
interpretation [6] in terms of the role of open strings in bosonic
D-branes, namely it corresponds to the decay of a D25-brane to give a
vacuum state in which only closed string states occur. This picture has
received convincing evidence in that the energy between the old and the
new vacua  agrees at low levels with the energy of the D25-brane [7,8]
and the open string states are in the zero cohomology class of  $Q$
when evaluated at the new vacuum [9]. 
\par
It has been realised that one could carry out a similar calculation for
the closed bosonic string, but little work has been done in this
direction as there has been no corresponding interpretation of the
solution and as closed bosonic string field theory is much more
complicated. In this paper we will argue that the appearance of
non-trivial  space-time derivatives in the open and closed string field
theories is a  feature that is   also  occurs in the
Landau theory of liquid-crystal transitions [10,11]. In this theory the
bilinear  term in the free energy has a momentum dependence that is not
simply given by
$p^\mu p_\mu$, but is a  more complicated function which possess a minimum
energy for a non-zero value of $p_\mu$. Further consideration of the
interaction terms implies that this corresponds to the crystalisation of
the liquid. 
\par
We will apply this
idea to the open and closed bosonic strings and find that for the open
string there appears to be no evidence of the appearance of this
mechanism, but for the closed bosonic string there exists a non-vanishing
value of the momentum that lowers the energy and so  spontaneously
breaking the  Lorentz and translation invariance of the theory. 
 We interpret this as meaning that the closed bosonic string theory 
undergoes a 
spontaneous compactification.  The possible relevance of the Landau
theory   of liquid-crystal transitions for   the closed bosonic string has
been discussed before [12].  
\par
Let us
begin by considering a simplified model that agrees with the lowest
level string field theory for the open and closed bosonic string for
appropriate values of its parameters. In particular, 
 we consider a scalar field
$\varphi$ which has the action 
$$A=\int d^D x\big( -{1\over 2} \partial_\mu \varphi \partial^\mu \varphi
+{b\over 2\alpha'}\varphi^2-\lambda\tilde \varphi^3\big)
\eqno(1.1)$$
where $\tilde \varphi=e^{\alpha'\gamma\partial^2}\varphi$ and 
$b,\lambda$ and $\gamma$ are dimensionless constants and $\alpha'$ is a
constant with the dimensions of ${\rm mass}^{-2}$. The potential for this 
action has an unstable minimum at the origin,  but a stable minimum at 
$<\varphi>={b\over 3\alpha'\lambda}$. After making the transformation to
the minimum by taking $\varphi=\hat \varphi +<\varphi>$, dropping the
$\hat {}$ and going to momentum space the action becomes 
$$A=-{1\over 2}\int {d^D p\over (2\pi)^D}\varphi(-p)D(p)\varphi(p)
$$
$$-\lambda \int {d^D r\over (2\pi)^D} \int {d^D q\over (2\pi)^D} \int {d^D
p\over (2\pi)^D}(2\pi)^D\delta^D (r+q+p)\varphi(r) \varphi(q) \varphi(p)
e^{-\alpha'\gamma (r^2+q^2+p^2)}\eqno(1.2)$$
where 
$$D(p)= {1\over \alpha'}(\alpha'p^2-{b}
+{2b}e^{-2\alpha'\gamma p^2}\big )
\eqno(1.3)$$
\par
We now wish to consider the energy for configurations which are time
independent, i.e. 
$\partial_0 \varphi=0$. As the action depends on an infinite number of
time derivatives the energy momentum tensor and so the energy have a
complicated expression, given in general in reference [14],  in terms of
the derivatives of
$\varphi$ and the differentials of  the Lagrangian by these quantities.
However, for time independent configurations the energy is just given by 
$E=-L$ where $L$ is the Lagrangian for time independent configurations. 
\par
We observe that the energy  has some features in common with the
energy used in the Landau theory of liquid-crystal phase transitions,
namely a kinetic energy term with a very non-trivial dependence on
momentum.  To analyse the energy we follow references [10,11] and first
examine 
 the kinetic term to find if there is a preferred minimum momentum, 
that is we look for the minimum of $D(p)$. In a usual field theory
$D(p)=p^2$ and the minimum is at $p^\mu=0$ and so the Lorentz and
translation symmetries are not broken. However, in our case we find that
a possible the minimum of $D(p)$ occurs at 
$$e^{2\alpha'\gamma p^2}=4\gamma b,\ {\rm or \ equivalently\  at},\ 
\alpha' p^2={1\over 2\gamma}\ln(4\gamma b)
\eqno(1.4)$$
As we are working in the mainly plus metric, $p^2=\underline p.\underline
p$ and so consequently we only find an actual  minimum if $4\gamma b>1$
assuming
$\gamma >0$. One easily verifies this is also just the condition for
$D(p)$ to decrease for small $p^2$, clearly indicating the presence of an
instability. For large
$p^2$
$D(p)$ goes to
$+\infty$. Hence if $4\gamma b>1$ we find that the state of lowest
energy has a non-vanishing momentum and the translation and Lorentz
symmetries are spontaneously broken. 
\par 
The direction of the momentum at
the minimum is not  determined by the kinetic energy term,  however, 
the cubic term contains a momentum delta function and so the momenta that
occur at the minimum must form a set of equilateral triangles. In the
Landau theory of the liquid-crystal transitions the field that occurs in
the free energy is the density which then forms a set of
oscillatory waves corresponding to the allowed momenta. The
intersecting coincident peaks of the density waves form a periodic array 
which is   interpreted as  a 
crystal. 
\par
The action of equation (1.1) is just that which occurs in the open and
closed bosonic string field theories if one works at the lowest
level and so keeps only the tachyon. The open bosonic string has been
extensively studied for the so called cubic open string formulation
which consists of only the quadratic term of references [3,4] and  a
term cubic  in string functional which was  given in  reference [4]. For
this theory at lowest level,  the parameters in the action of equation
(1.1) are given by [5]
$$\gamma=ln({3\sqrt 3\over 4})=0.2616,\ b=1,\ {\rm and}\  \lambda =({\sqrt
3\over 2})^7g_o
\eqno(1.5)$$
where $g_0$ is the open string coupling constant. 
One finds that $4\gamma b=1.0464$  and the corresponding possible minimum
momentum is given by 
$\alpha' p^2=0.0867$. It is difficult to argue  that these values are 
sufficiently in the interesting region to lead to the spontaneous
breaking of Lorentz and translation invariance. Indeed, it would seem more
likely that when higher orders are included the result will be
that the actual minimum occurs for 
$p^2=0$. In reference [5]  the kinetic energies for
various other fields are computed and one can carry out a similar
analysis. The result is that no convincing sign of a non-trivial minimum
momentum. 
\par
 Within the context of a  
non-polynomial string field theory for  the closed string, 
at the lowest level,  and keeping only the cubic term, reference [14]
found that the tachyon obeyed an action of the form of equation (1.1)
provided  the corresponding  parameters are given by  
$$\gamma ={1\over 2}ln({3\sqrt 3\over 4}),\ b=4, {\rm and}\  \lambda
=({ 3^8\over 2^{13}})g_c 
\eqno(1.6)$$
For the closed string one finds that 
$4\gamma b=2.0928$  and the corresponding $\alpha'
p^2=2.823=1.9962\sqrt 2$. It would seem conceivable that the actual
values of these quantities are 2 and $2.\sqrt 2$ respectively. 
 Hence, at least for the  lowest  order of approximation 
the closed bosonic string has a lowest energy state which  has
a non-vanishing momentum and so  spontaneously breaks translation and
Lorentz invariance. 
\par
The appearance of a minimum with non-vanishing momentum  solves a puzzle. 
Suppose that it had turned out that after shifting the scalar field to
the new vacuum that the energy was minimised by a  value   of the momentum
which vanished.  Assuming that this vacuum inherited to a solution
involving all the fields of the closed bosonic string theory in such a
way that translations and Lorentz symmetry  were also not broken then one
would have a solution 
 which preserved all these symmetries, but  had a lower energy than the 
vacuum where all fields and momenta vanished. However, this could not 
correspond to  a possible space-time filling brane solution, since, as
seen from the perspective of the closed string, such a solution  would
have positive energy with respect to the   original vacuum.  Although 
there should exist solutions of the closed bosonic string corresponding to
p-branes solutions which would break Lorentz and translation invariance 
these should also  have possess positive energy with respect to the zero
solution and so could not be identified with the vacua found above. 
\par
One can expect that the other fields of the closed bosonic
string also have kinetic terms whose lowest energy state is one with a
non-vanishing value of the momentum. Since one of these fields is the
graviton, which determines the underlying space-time, it is natural to
suppose that  the periodic nature of the vacuum state in space-time
actually corresponds to a compactification of space-time. The simple
calculation presented in this paper would have to be extended to higher
levels and so necessarily involve possible vacuum expectation values for 
tensorial fields before it became clear what compactifications would be
allowed. In the context of the open string,  vacua with non-vanishing 
expectation values for tensorial fields, but vanishing momentum, 
 have been considered [18] 
and it would be interesting to find an interpretation for them. 
\par
Corresponding to the spontaneous breaking of Lorentz and translation
symmetry we would expect the appearance of Goldstone bosons. These
would be restricted to propagate in the uncompactified space-time
directions and, due to the inverse Higgs effect,  their number would not
correspond to the number of broken generators. 
\par 
There is one very appealing interpretation for the
new vacuum solution  proposed in this paper.  The closed bosonic string
is a much simpler and more natural structure than  the closed
superstrings in ten dimensions and once a shift to a new vacuum can be
established it would appear to be a consistent theory. It would seem
unreasonable if such a tightly constrained theory was   not part of the
unified  picture, containing the ten dimensional string theories,  that
we now believe in. Indeed, sometime ago it was suggested that there could
exist a mechanism whereby the closed bosonic string could become the
superstring theories in ten dimensions. Although no such mechanism was
suggested, it has been shown [15] that there existed truncations of the
closed bosonic string spectra which were those of the ten dimensional
strings and,  more recently, this statement was extended to branes [16].
It is very natural to suppose that the new Lorentz and translation
breaking vacuum of the closed bosonic string found in this paper is the
missing mechanism that recovers the ten dimensional strings from the
closed twenty six dimensional bosonic string. Clearly, there
 remains much work to be done to establish the validity of the new
vacuum. It would be interesting to extend the calculation to include 
higher levels of the closed bosonic string and  although the method of
calculation given in reference [11] has good phenomological motivation,  it
would also be good to  give a more unified analysis of the energy
by treating the kinetic and higher order terms in the same way. However,
we note that the mechanism  is rather insensitive to the details of the
model and so one would expect it to be present also once these
corrections are included. 
\par
The existence of such a connection between the closed bosonic string and
the ten dimensional superstrings is consistent with the conjectures of
reference [17] where the former and the  latter, which are of type II,
were conjectured  to be invariant under  rank 27  and 11 Kac-Moody
algebras respectively.  Hence, although the closed bosonic string possess
no supersymmetry it  is thought, according to this conjecture, to be
invariant under a very powerful algebra that should determine many of its
properties. As a result, the vacuum found in this paper should break not
only Lorentz and translation symmetries, but also the $K_{27}$ algebra of
the closed bosonic string  and one may hope to find, at least  in the case
of breaking to the type II strings, that the  relevant vacuum
preserves the 
$E_{11}$ algebra. One encouraging sign is  that $K_{27}$ contains the
sub-algebra  
$E_{11}\oplus D_{16}$. While the first factor is clearly that required
as a residual symmetry, the second factor contains a $D_8$ sub-algebra 
which is required in the work of reference [15,16] to find the spectrum
of the superstrings in ten dimensions.  

\medskip
{\bf {References}}
\medskip
\item{[1]} M. Kaku and K. Kikkawa, {\it Field Theory of Relativistic
Strings. I. Trees}, Phys. Rev. D10 (1974) 1110; {\it Field Theory of
Relativistic Strings. II. Loops}, Phys. Rev. D10 (1974) 1823. 

\item{[2]}  A. Neveu and P. West, {\it  Gauge Covariant
Local Formulation of Bosonic Strings}; Nucl. Phys. {\bf B268} (1986) 125;
{\it The Interacting Gauge Covariant Bosonic String}, 
 Phys. Lett. {\bf 168B} (1985) 192, 
{\it Symmetries of the Interacting Gauge Covariant Bosonic String}, Nucl.
Phys. {\bf B278} (1986) 601; {\it String Lengths in Covariant String Field
Theory and  OSp(26,2/2)},  Nucl. Phys. {\bf B293} (1987) 226.

\item{[3]} A. Neveu, H. Nicolai and P. West, {\it New Symmetries and Ghost
Structure of Covariant String  Theories},  Phys. Lett. {\bf167B}
(1986) 307. 

\item{[4]} E. Witten, {\it  Non-Commutative Geometry and String Field
Theory}, Nucl. Phys. {\bf B268}  (1998) 253.

\item{[5]} V. Kostelecky and S. Samuel, {\it The Static Tachyon Potential
in the Open Bosonic string Theory}, Phys. Lett. {\bf 207B} (1988) 169,
{\it On a non-perturbative Vacuum for the Open bosonic String},  Nucl.
Phys. {\bf B336} (1990) 263. 

\item{[6]} A. Sen, {\it Stable Non-BPS Bound States of BPS D-branes}, JHEP
9808 (1998) 010, hep-th/9805019; {\it Descent Relations among Bosonic
D-branes}, hep-th/9902105,{\it  Universality of the Tachyon Potential},
JHEP9912 (1999) 027, hep-th/9911116. 

\item{[7]} A. Sen and B. Zwiebach, {\it Tachyon Condensation in String
Field theory}, hep-th/9912249. 

\item{[8]} N. Moeller and Washington Taylor, {\it Level Truncation and the
Tachyon in the open Bosonic String Field Theory}, hep-th/0002237. 

\item{[9]} I. Ellwood and Washington Taylor, {\it Open String Field Theory
without Open Strings}, hep-th/0103085. 

\item{[10]} For a review see P. W.  Anderson, {\it Basic Notions of
Condensed Matter Physics}, Benjamin/Cummings, 1984. 

\item{[11]} S. Alexander and J. McTague,
{\it Should all Crystals be bcc Landau Theory of Solidification and
Crystal Nucleation}, Phys. Rev. lett.  41 (1978) 702. 

\item{[12]} E. Rabinovici, {\it Introduction to String Ground State
construction},  CERN.4726/87; S. Elitzur, A. Forge and 
E. Rabinovici, {\it Some Global Aspects of String Compactification}, 
Nucl. Phys. B559 (1991) 581-610. 

\item{[13]} N. Moeller and B. Zwiebach, {\it Dynamics with Infinitely many
Time Derivatives and Rolling Tachyons}, hep-th/0207107

\item{[14]} V. Kostelecky and S. Samuel, {\it Collective Physics in
the Closed Bosonic String}, Phys. Rev D43, (1990) 1289. 

\item{[15]} A. Casher, F. Englert, H. Nicolai, A. Taormina, {\it
Consistent Superstrings as Solutions of the D=26 Bosonic String Theory}
Phys. Lett {\bf B162 } (1985) 121;
F. Englert, H. Nicolai, A. Schellekens, 
{\it Superstrings from $26$ dimensions},  
Nucl. Phys. {\bf B274} (1986) 315. 

\item{[16]} F. Englert, L. Houart and  A. Taormina, {\it   Brane Fusion in
the Bosonic and the Emergence of Fermionic Strings}, 
Phys. Lett {\bf JHEP 0108} (2001) 013, hep-th/0106235; 
A. Chattaraputi,  F. Englert, L. Houart and  A. Taormina, {\it The
Bosonic Mother of Fermionic D-branes },  hep-th/0207238. 

\item{[17]} P. West, {\it $E_{11}$ and M Theory}, Class. Quant. Grav. 18
(2001) 4443 , hep-th/0104081. 

\item{[18]} V. Kostelecky and S. Samuel, 
{\it Spontaneous Breaking of Lorentz Symmetry in String Theory}, 
Phys. Rev. {\bf D 39} (1989) 683; V. Kostelecky and R. Potting, {\it 
Expectation Values,Lorentz Invariance, and CPT in the Open Bosonic String}, 
hep-th/9605088.

\end